\documentclass[12pt]{iopart}
\usepackage{graphicx}
\begin{document}

\title[Onset of Magnetic Order in Transition Metal Oxides]
{Onset of Magnetic Order in Strongly-Correlated Systems from \emph{ab initio} 
Electronic Structure Calculations: Application to Transition Metal Oxides}

\author{I. D. Hughes$^1$, M. D\"ane$^{2,3}$, A. Ernst$^2$, W. Hergert$^3$, 
M. L\"{u}ders$^4$, J. B. Staunton$^1$, Z. Szotek$^4$ and W. M. Temmerman$^4$}
\address{$^1$ Department of Physics, University of Warwick, Coventry, CV4 7AL, UK}
\ead{j.b.staunton@warwick.ac.uk}
\address{$^2$ Max-Planck-Institut f\"ur Mikrostrukturphysik, Weinberg 2, D-06120 Halle, Germany}
\address{$^3$ Institut f\"ur Physik, Martin-Luther-Universit\"at Halle-Wittenberg,
Friedemann-Bach-Platz 6, D-06099 Halle, Germany}
\address{$^4$ Daresbury Laboratory, Daresbury, Warrington, WA4 4AD, UK}

\begin{abstract}
We describe an \emph{ab initio} theory of finite temperature magnetism in strongly-correlated
electron systems. The formalism is based on spin density functional theory, with a self-interaction
corrected local spin density approximation (SIC-LSDA). The self-interaction correction is implemented
locally, within the KKR multiple-scattering method. Thermally induced magnetic fluctuations are
treated using a mean-field `disordered local moment' (DLM) approach and at no stage is
there a fitting to an effective Heisenberg model. We apply the theory to the
3d transition metal oxides, where our calculations reproduce the experimental ordering tendencies,
as well as the qualitative trend in ordering temperatures. 
We find a large insulating gap in the paramagnetic state which hardly changes 
with the onset of magnetic order.
\end{abstract}

\submitto{\NJP}
\maketitle

\section{Introduction}
Many materials characterised by strong electron-electron correlations are of technological interest.
In the case of dilute magnetic semiconductors and the colossal magnetoresitive manganites, amongst others, 
this interest stems from their potential application in spintronic devices. As such, it is important
to understand the interplay between charge and spin degrees of freedom of the many interacting electrons 
in this class of materials. This requires a full quantum description of the electron spin. 
For low temperatures, a magnetic material has an electronic structure which has a fixed spin-polarisation, 
e.g. a uniform spin-polarisation for a ferromagnet or fixed sublattice spin polarisations for an antiferromagnet.  
With increasing temperature, spin fluctuations are induced which eventually destroy the long-range magnetic order 
and hence the overall spin polarization. These collective electron modes interact as the temperature increases, 
depending upon and affecting the underlying electronic structure. The implications can be explored by 
invoking a time scale separation between the fast electronic motions and the, much slower, spin fluctuations. 
For intermediate times, $\tau$, the spin orientations of electrons leaving an atomic site are sufficiently 
correlated with those arriving that the magnetisation, averaged over $\tau$, is nonzero. 
`Local-moments' are thus established. Although set up by the collective motion of the interacting electrons 
these can be described as classical spin like variables,$\{\mathbf{\hat{e}}_i\}$, provided the temperature is not 
so low that the dynamics of the spin fluctuations become important. So, for most finite temperatures, 
ensemble averages over the static orientational configurations of the local moments determine the magnetic properties 
of a system. In the so-called disordered local moment (DLM) approach \cite{Hubbard}, these
averages are carried out using a mean-field technique.

A first-principles implementation of the DLM method \cite{Gyorffy}, based on density functional theory (DFT) in the 
local spin density approximation (LSDA) and a multiple scattering effective medium method to handle the local 
moment disorder, has proved to be extremely successful at describing the magnetic properties of many transition
metal systems ~\cite{Staunton,Staunton:2,Razee}. In these systems, however, charge correlations are weak and the electrons 
are fully itinerant. This is certainly not the case in strongly-correlated systems, such as the transition 
metal oxides (TMO), which often contain highly localised electron states for which the LSDA fails. 

The self-interaction corrected local spin density approximation (SIC-LSDA) \cite{Perdew}, can account for 
correlation effects such that the ground state properties of the TMOs are well reproduced ~\cite{Svane,Szotek}. 
Recent papers by Lueders et al.~\cite{Lueders} and Daene et al.~\cite{Daene} set out how to implement this 
approach using a multiple scattering Kohn-Korringa-Rostoker (KKR) method. This implementation of the SIC, 
the so-called local SIC (LSIC) \cite{Lueders} offers an immediate generalisation to disordered systems 
and opens up the possibility of implementing the DLM picture of magnetism within such a SIC based electronic 
structure scheme. 

Indeed recently~\cite{Hughes}, we showed how such a DLM-SIC approach provides an accurate description of 
finite temperature magnetism in the heavy rare earths. In these systems, the magnetic moments arise from 
highly localised 4f states and are coupled via an indirect RKKY exchange mechanism \cite{Kasuya}, mediated 
by the sd conduction electrons. In this paper we apply the DLM-SIC to the TMOs, where a superexchange 
mechanism, mediated by the oxygen ions, operates between local moments set up by the 3d states of the 
transition metal ions. We investigate the spin fluctuations that characterise the paramagnetic state of 
these systems above their magnetic transition temperatures, gaining information about the type of 
ordering that is likely to occur as the temperature is lowered through a phase transition. This finite 
temperature study is complementary to calculations of the ground state energies of the same systems in 
different magnetic states at $T=0$K carried out by several groups~\cite{Svane,Szotek,Diemo}. The work by 
Daene et al.~\cite{Daene} is particularly relevant. 

The paper is organised as follows. In section \ref{formalism} we outline the DLM formalism and its first-principles
SIC-LSDA implementation within the KKR method. We examine in detail the electronic structure
of MnO in the paramagnetic state in section \ref{MnO}, and, by evaluating the paramagnetic spin susceptibility, investigate
the presence of any underlying ordering tendencies. We extend our study to the whole TMO series in section \ref{TMO}, where
we find good agreement with experiment, with the notable exception of the N\`{e}el temperature of NiO, 
which we find to be too small by a third. The paper is summarised in section \ref{conclusion}, where we include some 
discussions as to what additional aspects, not accounted for in our present formalism, may be needed in order
to capture fully the physics of NiO. 

\section{Formalism}
\label{formalism}
We start with a key assumption of a time scale separation between that associated with 
fast electronic motions, i.e.\ the electron hopping time scale, and that characteristic
of, typically much slower, spin fluctuations. At an intermediate time scale well defined
moments exist at all lattice sites, the orientations of which we describe 
using a set of unit vectors, $\{\mathbf{\hat{e}}_i\}$. The local moment phase 
space specified by $\{\mathbf{\hat{e}}_i\}$ is assumed to be ergodic and hence long 
time averages can be replaced by ensemble averages. These averages use the Gibbsian measure
\begin{equation}
P (\{\mathbf{\hat{e}}_i\}) = Z^{-1}\exp[-\beta \Omega (\{\mathbf{\hat{e}}_i\}) ],
\end{equation}
where the partition function, $Z$, is given by
\begin{equation}
Z =  \prod_j \int d \mathbf{\hat{e}}_j \exp [-\beta \Omega (\{\mathbf{\hat{e}}_i\}) ]
\end{equation}
and $\beta=(k_{B}T)^{-1}$. $\Omega (\{\mathbf{\hat{e}}_i\})$ is a `generalised' grand potential, the term `generalised'
meaning that $\Omega (\{\mathbf{\hat{e}}_i\})$ is not associated with a thermal equilibrium state.

In the DLM approach~\cite{Gyorffy, Staunton:2} a mean-field approximation for $\Omega (\{\mathbf{\hat{e}}_i\})$ 
is constructed by expanding it about a single-site reference spin Hamiltonian, 
$\Omega_{0} (\{\mathbf{\hat{e}}_i\})=-\sum_{i} \mathbf{h}_i \mathbf{\cdot} \mathbf{\hat{e}}_i$.
Here, the parameters $\mathbf{h}_i$ play the role of a Weiss field and are determined using a Feynman variational approach 
\cite{Feynman}, whereby the free energy of the system, $F=-\beta^{-1}\mathrm{log}Z$, is minimised. This free energy 
takes into account both the entropy associated with transverse spin fluctuations and also the production of electron-hole 
pairs associated with Stoner excitations. 

The probability function, $P_{0}$, associated with $\Omega_{0}$ can be written as a product of single-site measures:
\begin{equation}
P_{0}(\{\mathbf{\hat{e}}_i\})=\prod_{i} P_{i}(\mathbf{\hat{e}}_i),
\label{prob0}
\end{equation}
where
\begin{equation}
P_{i}(\mathbf{\hat{e}}_i)=Z_{i}^{-1}\exp[\beta \mathbf{h}_i \mathbf{\cdot} \mathbf{\hat{e}}_i]
\label{prob}
\end{equation}
and
\begin{equation}
Z_{i} =  \int d \mathbf{\hat{e}}_i \exp [\beta \mathbf{h}_i \mathbf{\cdot} \mathbf{\hat{e}}_i].
\label{partition}
\end{equation}
Specifying these single-site probabilities, $P_{i}$, means that a class of mean-field theories, 
developed originally for the study of substitutionally disordered alloys, becomes available to us to treat
the local moment disorder. In particular, we deploy the coherent potential approximation (CPA) \cite{Soven}, 
which is known to be the best single-site approximation. Here, an effective medium is specified.
The motion of an electron through this medium approximates, on the average, the motion of an electron in the 
disordered lattice. This effective medium is determined self-consistently and in the context of multiple-scattering theory, 
i.e.\ the so-called KKR-CPA \cite{Stocks,Faulkner}, the self-consistency condition states that there should be 
no further scattering of an electron, on the average, when a single site in the effective medium is replaced by 
one of the constituent `alloy' potentials. For the local moment disorder, this condition reads mathematically as
\begin{equation}
\int d \mathbf{\hat{e}}_i P_{i}(\mathbf{\hat{e}}_i) \tilde{D}_{i}(\mathbf{\hat{e}}_i)=\tilde{1},
\label{CPA:DLM}
\end{equation}
where quantities with a tilde superscript (\~{}) are $2 \times 2$ matrices in spin
space and $\tilde{D}_{i}(\mathbf{\hat{e}}_i)$ are the so-called \emph{CPA projectors}, 
defined by (in $L (\equiv l,m)$ representation) 
\begin{equation}
\tilde{D}_{i}(\mathbf{\hat{e}}_i)=[\tilde{1}+((\tilde{t}(\mathbf{\hat{e}}_i))^{-1}-
(\tilde{t}^{c}_{i}) ^{-1})\tilde{\tau}^{c,00}]^{-1}.
\label{projector}
\end{equation}
Here, the single-site matrix $\tilde{t}(\mathbf{\hat{e}}_i)$  describes the scattering from a 
site with local moment orientated in the direction $\mathbf{\hat{e}}_i$ such that 
\begin{equation}
\tilde{t}_{i}=\frac{1}{2}(t_{+}+t_{-})\tilde{1}+\frac{1}{2}(t_{+}-t_{-})\tilde{\mathbf{\sigma}} \cdot \mathbf{\hat{e}}_i,
\label{t}
\end{equation}
where $\tilde{\sigma}_{x}$, $\tilde{\sigma}_{y}$ and $\tilde{\sigma}_{z}$ are the three 
Pauli spin matrices defined according to the global z-axis.
In the local reference frame, where the z-axis is aligned with $\mathbf{\hat{e}}_i$, we evaluate the 
matrices $t_{+}$/$t_{-}$, representing the scattering of an electron with spin parallel/antiparallel to the 
local moment direction $\mathbf{\hat{e}}_i$. These matrices are calculated according to 
\begin{equation}
t_{+(-)L}(\epsilon)=-\frac{1}{\sqrt\epsilon}\sin\delta_{+(-)L}(\epsilon)e^{i\delta_{+(-)L}(\epsilon)},
\end{equation}
where the phaseshifts $\delta_{L}(\epsilon)$ are computed using effective DFT potentials. These effective 
potentials, $v_{+}$ and $v_{-}$, differ on account of the `local exchange splitting'~\cite{Gyorffy}, which is the 
cause of the local moment formation. Unlike the conventional LSDA implementation,
the potentials $v_{+}/v_{-}$ are orbital dependent in our new SIC-LSDA approach. This dependency comes about
by our SI-correcting certain $L$ channels, the details of which can be found in reference~\cite{Daene}. 
Importantly, the SI-corrected channels of $v_{+}$ and $v_{-}$ may differ. Indeed the valence channels to which
we apply the self-interaction correction are those with a resonant phase shift~\cite{Lueders,Daene}. Such 
resonant behaviour is characteristic of well localised electron states, which will establish quasi-atomic like moments.
Through the influence they exert on the electron motions, these moments will be reinforced by the spins of 
more itinerant-like electrons. It thus follows that resonant states will tend to define the local moment orientation 
and, as such, we expect to SI-correct a greater number of channels of $v_{+}$ than we do for $v_{-}$.
 
$\tilde{t}^{c}_{i}$ describes a site of the CPA effective medium. The scattering-path matrix, $\underline{\tilde{\tau}}$ 
(where underlined matrices are in site representation), is related to the single-site scattering matrices $\tilde{t}$ via
\begin{equation}
\underline{\tilde{\tau}} (\epsilon)=[\underline{\tilde{t}}^{-1}(\epsilon)-\underline{g}(\epsilon) \tilde{1}]^{-1},
\end{equation}
where $\epsilon$ is a complex energy and $\underline{g}$ is the structural Green's function, which describes
the free propagation of an electron between scattering centres. 

In the paramagnetic regime, $\tilde{t}_c = t_c \tilde{1}$ and $\tilde{\tau}^{c,00}= \tau^{c,00}\tilde{1}$, the local 
moments have no preferred orientation and the $P_{i}$'s become site independent. Moreover using Eq.~\ref{t} we find
\begin{equation}
\tilde{D}^{0}_{i}=\frac{1}{2}(D^{0}_{+}+D^{0}_{-})\tilde{1}+\frac{1}{2}(D^{0}_{+}-D^{0}_{-})
\mathbf{\tilde{\sigma}} \cdot \mathbf{\hat{e}}_i
\label{D}
\end{equation}
where 
\begin{equation}
D_{+(-)}^{0}=[1+[(t_{+(-)}^{-1}-(t^{c})^{-1}]\tau^{c,00}]^{-1}.
\end{equation}
The superscript $0$ signifies that the CPA projector is evaluated in the paramagnetic state.
Substituting $P_{i}(\mathbf{\hat{e}}_i)=P^{0}=\frac{1}{4\pi}$ into Eq. \ref{CPA:DLM} we obtain 
\begin{equation}
\frac{1}{4\pi} \int d \mathbf{\hat{e}}_i \tilde{D}^{0}_{i}(\mathbf{\hat{e}}_i)=\tilde{1},
\label{CPA:para}
\end{equation}
which becomes, on carrying out the integration,
\begin{equation}
\frac{1}{2}D_{+}^{0}+\frac{1}{2}D_{-}^{0}=1.
\label{CPA:Ising}
\end{equation}

Equation \ref{CPA:Ising} is evidently just the CPA equation for a system with 50\% of moments pointing 
`up' and 50\% pointing `down', i.e. an Ising-like system. The electronic structure problem is thus reduced 
to that of an equiatomic binary alloy, where the two `alloy' components have anti-parallel local moments. 
Treating this `alloy' problem with the KKR-CPA, in conjunction with the LSIC charge self-consistency 
procedure outlined in references~\cite{Lueders, Daene}, we arrive at a fully self-consistent LSIC-CPA 
description of the DLM paramagnetic state. 

It should be noted that the equivalence of the DLM electronic structure problem to that of an Ising-like system is purely
a consequence of the symmetry of the paramagnetic state, and is not the result of our imposing any restriction
on the moment directions. Indeed, in the formalism for the paramagnetic spin susceptibility, which we outline now, 
we maintain and consider the full 3D orientational freedom of the moments. 

Within the DLM method the magnetisation at a site $i$, $\mathbf{M}_i$, is given by 
$\mathbf{M}_i=\mu\mathbf{m}_i$, where $\mathbf{m}_i=\int \mathbf{d \hat{e}}_{i} \ P_{i}(\mathbf{\hat{e}}_{i}) \mathbf{\hat{e}}_{i}$ and $\mu$ is the local moment magnitude, determined self-consistently. 
In the paramagnetic regime, where $P_{i}$ is independent of $\mathbf{\hat{e}}_{i}$, $\mathbf{M}_i=\mathbf{0}$.
Using a perturbative approach, we investigate the onset of magnetic order, where $\mathbf{M}_i=\mathbf{0}$ becomes finite.
In particular, we consider the response of the paramagnetic state to the application of an external, site-dependent 
magnetic field. Focusing on the dominant response of the system to line up the moments with the applied field, we 
obtain the following expression for the  static spin susceptibility:
\begin{equation}
\chi_{ij}=
\frac{\beta}{3}\mu_{i}^{2}\delta_{ij}+\frac{\beta}{3}\sum_{k}S^{(2)}_{ik}\chi_{kj},
\label{susc}
\end{equation}
where $S^{(2)}$ is the direct correlation function for the local moments, defined by
\begin{equation}
S^{(2)}_{ik}=-\frac{\partial^{2} <\Omega>}{\partial m_{i} \partial m_{k}}.
\end{equation}

In the paramagnetic state, $S^{(2)}_{ik}$ depends only on the vector
difference between the positions of sites $i$ and $k$. A lattice Fourier transform
can hence be taken of Eq.~\ref{susc}, giving
\begin{equation}
\chi\left(\mathbf{q}\right)=\frac{1}{3}\beta\mu^{2}[1-\frac{1}{3}\beta S^{(2)}\left(\mathbf{q}\right)]^{-1}.
\label{susc:LFT}
\end{equation}
An expression for $S^{(2)}\left(\mathbf{q}\right)$, involving scattering quantities obtained from the electronic
structure of the paramagnetic state, can be found in reference \cite{Staunton:susc}. 

By investigating the wavevector dependence of the susceptibility we gain information about the
spin fluctuations that characterise the high temperature paramagnetic state. Conversely most
first-principles calculations for the TMOs have, up to now, concentrated on the groundstate.
These $T=0$ calculations have demonstrated the importance of including strong electron
correlation effects. In particular for conventional LDA calculations, 
in which such effects are neglected, the size of the groundstate magnetic moments are found 
to be substantially underestimated. When the effects of strong Coulomb repulsion between electrons 
occupying the partially filled d states are taken into account, such as through the self-interaction 
correction \cite{Svane,Szotek,Diemo} or LDA+U method \cite{Dobysheva}, the resulting moments are found to be in
much better agreement with experiment. 

In some respects, the incorporation of correlation effects is even more important when
dealing with finite temperatures. In particular, it is essential to describe
the localised nature of the d states in order that local moments survive into the paramagnetic phase. 
Indeed, when strong correlation effects were neglected in our DLM calculations we were not
able to stabilise a local moment for NiO. For the other TMOs a local moment could be stabilised but, 
contrary to experimental observations, a large magnetovolume effect was exhibited.

Recently, using the LDA+DMFT method \cite{Anisimov}, it was shown how, by taking suitable 
account of strong electron-electron correlations, a reasonable description 
of the electronic spectrum of NiO in its paramagnetic state can be obtained \cite{Ren}. 
In another LDA+DMFT study \cite{Wan}, in addition to NiO, calculations for other TMOs have also been performed.
Our DLM-SIC investigation of the TMO series, which we go on to describe now, focuses on the onset of magnetic 
%order and can be considered complementary to these studies. 
order and can be considered equivalent to these studies, but without quantum fluctuations. 
In particular, we concentrate on
the energy scales associated with magnetic (spin) fluctuations, enabling us to obtain
%estimates of the N\'{e}el temperatures.
estimates of the N\'{e}el temperatures, 
and we investigate the importance of the dynamical quantum fluctuations.
\section{MnO}
\label{MnO}
In this section we discuss in detail our calculations for MnO. Using the LSIC-CPA approach
outlined in section \ref{formalism}, we perform an electronic structure calculation for the paramagnetic
state. We use the atomic sphere approximation (ASA) with a unit cell that has, in addition
to a manganese and an oxygen ion, two empty spheres so as to obtain a better representation of the charge 
density and space filling. Daene et al.~\cite{Daene} compare the energies of the transition metal oxides for 
specific magnetically ordered states at $T=0$K between different SIC configurations corresponding to different 
numbers of localised states. The configuration of lowest energy is determined by the balance between localisation 
(self-interaction) and band formation (hybridisation) energy. 
We follow the same procedure here but for the high temperature paramagnetic state.
For MnO, we find the energy to be minimised when all Mn d states with spins parallel to the local moment direction 
are SI-corrected, with none corrected in the opposite spin channel. This picture is in keeping with Hund's first rule.

\begin{figure}
\begin{centering}
\input{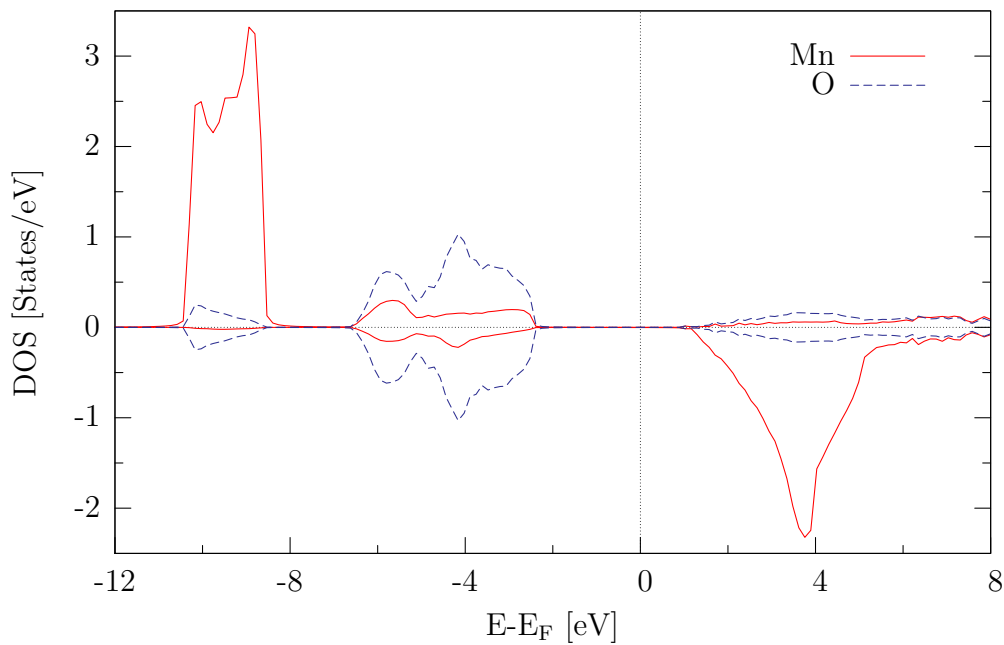}
\caption{The local density of states (DOS) for MnO in its paramagnetic (DLM) state on Mn (full line) and O sites 
(dashed). The upper (lower) panel shows the DOS associated with electrons with spins parallel (anti-parallel) to 
the local moment on the site.
Note that a sizeable gap persists in the paramagnetic state.}
\label{DOS}
\end{centering}
\end{figure}

In Figure \ref{DOS} we show the local DOS, where an exchange-splitting is evident. 
Of course, when an average is taken over all spin orientations, the electronic structure does not have an 
overall spin-polarisation. Nevertheless, it is possible to identify such `local exchange splitting' experimentally,
using photoemission \cite{Kisker} and inverse photoemission \cite{Kirschner} techniques. 
The local moment obtained for the Mn sites in the DLM paramagnetic state differs from that in the
groundstate by $\approx 0.03\mu_{\mathrm{B}}$ (see Table \ref{comparison}). Such a small change between the
ordered (zero temperature) and disordered (high temperature) states justifies our DLM picture, where the 
magnitude of the `local exchange splitting' is independent of the orientational configurations of the moments.
Furthermore this feature causes the spin-summed electronic structure to show little difference between the paramagnetic state above $T_N$ and the magnetically ordered ground state. Notably a sizeable band gap at the Fermi energy persists into the paramagnetic phase.

\begin{figure}
\begin{centering}
\includegraphics[scale=1.2]{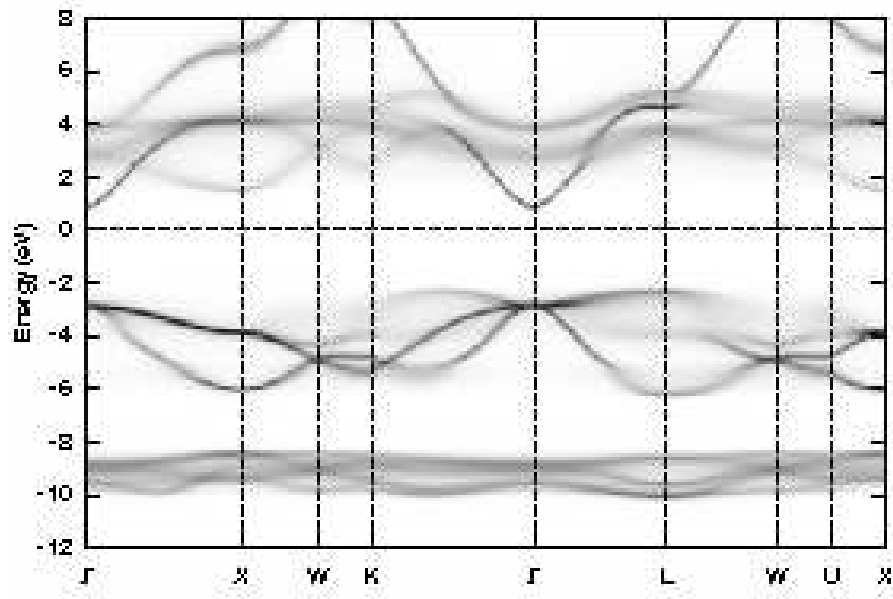}
\caption{The electronic structure for MnO in its paramagnetic (DLM) state along symmetry directions. 
The loci of the peaks of the Bloch spectral function are displayed and the shading shows the broadening of these 
quasi-particle peaks caused by the spin fluctuation disorder. }
\label{BSF}
\end{centering}
\end{figure}

Figure \ref{BSF} shows the electronic structure of the paramagnetic (DLM) state in more detail and 
demonstrates the wave-vector, ${\bf k}$, dependence of the local exchange splitting. 
This figure is constructed from calculations of the Bloch spectral function, $A_B ({\bf k}, E)$. For 
non-interacting electrons in an ordered system at $T=0$K, $A_B ({\bf k}, E)$ comprises a set of Dirac 
delta functions which trace out the electronic band structure. When electron interaction, finite 
temperature or disorder effects are included the spectral function is a set of broadened peaks 
describing quasi-particle excitations. The broadening of the peaks shown in Figure \ref{BSF} is a 
consequence of the local moment disorder in the paramagnetic state~\cite{Staunton:3}. 
%By looking at the positions of the peaks, which are well defined even after broadening, we 
%trace out an effective band structure, from which we determine an effective band gap value. 
%For MnO this gap value is ???$\mu_{\mathrm{B}}$, which is close to the band gap obtained in the groundstate 
%calculations of \cite{Daene} (see table \ref{comparison}).

\begin{figure}
\begin{centering}
\includegraphics[scale=0.75]{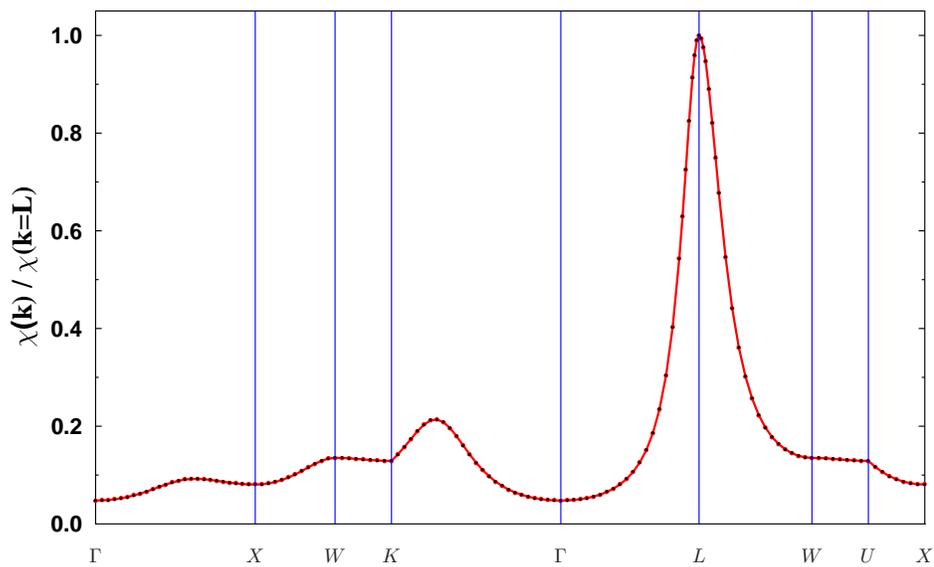}
\caption{Paramagnetic spin susceptibility of MnO, as a function of wavevector $\mathbf{k}$.}
\label{Fig:susc}
\end{centering}
\end{figure}

In figure \ref{Fig:susc} we present the results of our paramagnetic spin susceptibility calculations for MnO. These show 
the paramagnetic state to be dominated by spin fluctuations with wavevector $\mathbf{q}_{\mathrm{max}}
=(0.5,0.5,0.5)$ (in units of $2\pi/$(lattice constant)). This indicates that the system will order into an antiferromagnetic type II (AF2) structure, 
where moments within a $<$111$>$ layer are aligned but are antiparallel in successive layers. 
This concurs with the experimentally observed groundstate of this system and also 
$T=0$K calculations \cite{Svane,Szotek,Daene}, where the most stable structure was determined by 
comparing the total energies of different magnetic configurations.

We examine the temperature dependence of the static spin susceptibility $\chi(\mathbf{q}_{\mathrm{max}})$, in particular 
looking for a divergence which would signify that the paramagnetic state becomes unstable with respect
to the formation of a spin density wave, characterised by the wavevector $\mathbf{q}_{\mathrm{max}}$. 
For the theoretical (experimental) lattice parameters, such a divergence occurs at 102K (103K).
This mean field estimate of the N\'{e}el temperature is in good agreement with the experimental value of 118K
(see Table \ref{comparison}). 

\section{TMO series}
\label{TMO}
In this section we extend our study to the other TMOs which also occur in the cubic rocksalt structure above the N\'{e}el temperature, namely FeO, CoO and NiO. 
We find the energies to be minimised when all TM d states with spins parallel to the local moment direction are 
SI-corrected  together with one, two and three $t_{2g}$ d-states corrected in the opposite spin channel for FeO, CoO and 
NiO respectively. This is again in accord with Hund's rule. The magnetic moments obtained are listed in Table 
\ref{comparison} and are found to agree closely with their groundstate values. This concurs with experimental data for 
NiO \cite{Brandow}, where the local moment size remains essentially unchanged as the N\'{e}el temperature is passed 
through. Indeed, more generally, the transition between the magnetically ordered and disordered phases is known to 
have little effect on the valence band photoemission spectra \cite{Tjernberg},
and this is reflected in our spin-summed density of states which hardly changes
between the paramagnetic state above $T_N$ and the magnetically ordered ground state at $T=0$K.
The insulating gaps of the paramagnetic DLM states are also very close in magnitude to those found in our calculations of the magnetically
ordered states \cite{Daene}. We find the gaps' sizes, in eV,  of the paramagnetic states to be 3.3 (3.7) for MnO (see Figs.\ref{DOS},\ref{BSF}), 3.5 (2.5) for FeO, 2.8 (2.4) for CoO and 3.8 (4.1) for NiO, where the experimental values shown in brackets are taken from the summary in ref.~\cite{Svane} and also ~\cite{Bong}  

Our paramagnetic susceptibility calculations indicate that, like MnO, the other members of the TMO series have
a tendency to order into the AF2 structure. The temperatures at which we predict this ordering to 
occur are listed in Table \ref{comparison}, where we find good agreement with experiment, with
the exception of NiO where we underestimate the temperature by about a third. Interestingly,
the Heisenberg mean-field results from a recent investigation of the TMOs \cite{Harrison}, 
where a Hartree-Fock treatment of the electrons was used, show a similar discrepancy
for the N\'{e}el temperature of NiO. 
Also, recent quasiparticle self-consistent {\em GW} \cite{Kotani} results for MnO and NiO show similar behaviour.
This suggests that some additional physics, not at work
in the other transition metal oxides, may be of relevance to the determination of the ordering temperatures of NiO.
We return to discussing this point in section \ref{conclusion}. 
In Table \ref{comparison} we also give the N\'{e}el temperatures obtained from a muffin-tin (MT), as opposed to an ASA,
implementation of our electronic structure scheme. Qualitatively, the trend in temperatures is the same, although
the MT implementation gives generally better agreement with experiment. 

\begin{table}
\caption{Local magnetic moments and N\'{e}el temperatures for the series
of transition metal oxides, obtained from our DLM-SIC calculations. The magnetic moments
are compared to those obtained in reference~\cite{Daene} for the groundstate (AF2) configuration. 
Values not enclosed (enclosed) in round brackets were calculated using theoretical (experimental) lattice parameters.
The variation of the N\'{e}el temperatures with respect to changes in the lattice spacing is also given.} 
\begin{tabular*}{\textwidth}{@{}l*{5}{@{\extracolsep{0pt plus12pt}}c}}
\br                              
& \multicolumn{4}{c} {\bf Compound}\cr
& MnO&FeO&CoO&NiO\cr 
\mr
%\multicolumn{4}{l} {\bf Band gap [eV]} & \cr
%DLM         &???       &???       &???       &???\cr
%AF2         &3.29      &3.40      &2.76      &3.44\cr
%Expt. (AF2) &3.6-3.8$^{\mathrm{a}}$ &2.4$^{\mathrm{b}}$,2.5$^{\mathrm{c}}$ 
%&2.4$^{\mathrm{a}}$ &4.3$^{\mathrm{a}}$,4.0$^{\mathrm{a}}$,4.3$^{\mathrm{d}}$\cr
%\mr
\multicolumn{4}{l}{\bf Local magnetic moment on TM [$\mu_{\mathrm{B}}$]} & \cr
DLM         &4.63(4.63)&3.69(3.68)&2.71(2.71)&1.72(1.71)\cr    
AF2         &4.60(4.60)&3.68(3.66)&2.69(2.68)&1.68(1.67)\cr
Expt. (AF2) &4.79$^{\mathrm{a}}$,4.58$^{\mathrm{a}}$ &3.32$^{\mathrm{a}}$  
&3.35$^{\mathrm{b}}$,3.8$^{\mathrm{a}}$  &1.77$^{\mathrm{a}}$,1.64$^{\mathrm{a}}$,1.90$^{\mathrm{a}}$\cr
\mr
\multicolumn{4}{l} {\bf  N\'{e}el temperature [K]} & \cr
Theory(ASA)&102(103)&148(170)&228(250)&344(367)     \cr
Theory(MT) &   (129)&   (203)&   (283)&   (383)\cr
Expt.$^{\mathrm{c}}$ &118&198&291&525\cr
\mr
\multicolumn{4}{l} {\bf  $\frac{\partial T_{\mathrm{N}}}{\partial \mathrm{lat}}$ [K/a.u.]} & \cr
& -225 & -270 & -318 & -376\cr 
\br
\end{tabular*}
$^{\mathrm{a}}$Taken from ref.~\cite{Svane}, for detailed references see references therein.\\ 
%$^{\mathrm{b}}$Reference \cite{Bowen}.\\
%$^{\mathrm{c}}$Reference \cite{Kim}.\\
%$^{\mathrm{d}}$Reference \cite{Sawatzky}.\\
$^{\mathrm{b}}$Reference \cite{Khan}.\\
$^{\mathrm{c}}$Reference \cite{Kittel}.
\label{comparison}
\end{table}

\section{Conclusion}

\label{conclusion}
We have used a locally self-interaction corrected implementation of the disordered local moment 
model to study the onset of magnetic order in MnO, FeO, CoO and NiO. Specifically,
by taking into account the strong intrasite Coulomb correlations between the 3d electrons through SIC, we 
obtained an accurate first-principles, finite temperature description of the paramagnetic state. 
We found, in agreement with recent DMFT studies of Ren \emph{et al} \cite{Ren} and Wan \emph{et al} \cite{Wan},
that the paramagnetic state from our DLM-SIC study is characterized by a large insulating gap
(see Figs. \ref{DOS} and \ref{BSF} for MnO). We conclude that
this gap is the consequence of the strong electronic correlations in the paramagnetic
state which the DLM-SIC describes particularly well
without the need of including dynamical fluctuations.
Also the local moment obtained for the transition metal sites in the DLM-SIC paramagnetic state differs little from that in the ground state (e.g. by $\approx 0.03\mu_{\mathrm{B}}$ for MnO). 
We see also from the present work that the magnetic order in the AF2 ground state has no
significant influence on the spin-summed electronic structure of NiO.
This is consistent with our previous results that the band gap of AF1, AF2 and F magnetically ordered NiO differ only slightly\cite{Diemo} mostly on account of changes in hybridization.
With the successful application of the DLM-SIC to the transition metal oxides we have
been able to demonstrate that it is a generalized framework, capable
of computing finite temperature magnetic interactions for correlated systems as
well as the
itinerant systems\cite{Gyorffy,Staunton:2}, where the SIC reduces to LSDA.

The underlying magnetic ordering tendencies that our DLM-SIC study revealed were found to be in agreement with
the experimentally observed magnetic groundstates. The corresponding N\'eel temperatures were, 
with the notable exception of NiO, also in good agreement with experiment. 
In order to begin to understand our theoretical underestimate of the N\'eel temperature
of NiO, it is informative to look at the N\'{e}el temperatures of each of the TMOs as a function of lattice
spacing. We found the temperatures to vary approximately linearly with lattice spacing
and, as shown in Table \ref{comparison}, this dependency becomes more pronounced as the TMO series is crossed. 
This in turn implies that, as the series is crossed, the N\'{e}el temperatures become more sensitive to the 
underlying electronic structure. In particular, the hybridisation between the strongly-correlated d states and 
the exchange-mediating oxygen (s and p) sites becomes more delicate. This suggests that some
of the error in the N\'eel temperature might be due to the incorrect positioning of these states. 
Indeed, although the LSIC does a much better job than the LDA at reproducing 
the insulating gaps of TMOs,
%the experimental 
%spectra of the TMOs, significant errors still exist. In particular, the groundstate 
the LSIC band gaps
reported in \cite{Daene} and here, although qualitatively correct, are in disagreement with experiment by up to 40\% (FeO).
In the recent LDA+DMFT study of Ren \emph{et al} \cite{Ren}, the band gap of NiO was 
reproduced with very high accuracy, of course subject to an appropriate choice of the Hubbard U parameter. 
In that investigation a dynamic self-energy was used to incorporate 
correlation effects in contrast to the static self-energy used in the SIC approach. Our DLM 
treatment of essentially transverse magnetic fluctuations can also be considered as the static limit of some, as 
yet undeveloped, dynamical mean field theory of spin fluctuations. Since, however, we deal with temperatures
where the dynamics of transverse spin fluctuations should not be important, our static treatment 
of these electronic degrees of freedom is a reasonable approximation. 
The effects of the faster dynamical charge correlations and longitudinal spin fluctuations, on the other hand,
may be significant even at high temperatures and a more sophisticated account of these  
might therefore improve our estimate of the N\'eel temperature in due course. 

In this paper we have focussed on the paramagnetic states of the TM-oxides. Here the crystal structure is that of cubic rock salt. It has been well-documented \cite{Jauch2,Roth} how the establishment of AF2 magnetic order prompts a distortion of the crystalline lattices in these materials below $T_N$. By including relativistic effects, such as spin-orbit coupling, into our description of the TMOs in principal it should be possible to deduce how these magnetostrictive effects arise at $T_N$ as the symmetry is broken. The inclusion of spin-orbit coupling would also determine how the magnetic moments orient with respect to the crystal lattice as magnetic order develops.~\cite{Ressouche, Vijay}
This fundamental approximation made in our investigation of the neglect of spin-orbit coupling needs some further comment. For
transition metals this is a good approximation since crystal field effects quench the 3d orbital moments \cite{Cox}. 
In TMOs, however, strong Coulomb correlations can lead to a reduction of these crystal field effects
and hence a preservation of the orbital moment. Indeed, the presence of an orbital moment in CoO
has long been established \cite{Terakura}. Recent experimental \cite{Neuback} and theoretical \cite{Kwon}
work has suggested that a significant orbital moment is also exhibited by NiO, with an orbital to spin angular momentum ratio 
as high as L/2S=0.17 \cite{Neuback}. There are important implications associated
with the presence of an orbital moment, in particular with regards to which orbitals are occupied. More specifically, 
for a doublet groundstate the orbital angular momentum
is completely quenched \cite{Abragam} and hence the presence of an orbital moment means that the d states
cannot be in a pure $(t_{2g})^{6}(e_{g})^{2}$ configuration. A recent experimental study of NiO \cite{Jauch} 
has suggested that, in the groundstate, there is a fractional occupation of orbitals by unpaired spin electrons.
Due to a very slight rhombohedral distortion, brought
about by antiferromagnetic ordering, the symmetry changes from cubic to rhombohedral below the N\'eel 
temperature. As a result, the $t_{2g}$ orbital splits into a single $a_{g}$ and doubly degenerate $e_{g}$ level, 
with the original $e_{g}$ level retaining its symmetry characteristics but renamed $e'_{g}$. 
In reference \cite{Jauch} it was reported that the $a_{g}$ level has a partial filling of 1.69, giving rise to an 
orbital moment $\mu_{L}=0.31\mu_{B}$. In our calculations, although we allow the energetics to determine the 
orbital configurations, we are restricted to integer occupancies of the SI-corrected, localised, orbitals. In order
to consider fractional occupancies it is feasible to use the CPA
to describe a random distribution of orbital configurations, analogous to our treatment of random moment orientations 
described in section \ref{formalism}. Through careful choice of the
probabilities associated with the different orbital configurations, various
fractional occupancies can be mimicked to be followed by an investigation of orbital ordering in the TMOs. 
This type of study is being planned and holds promise for a further contribution to the understanding of these materials.

\section*{References}

\bibliographystyle{unsrt.bst}
%\bibliography{references}

\end{document}